# SNTL-NTU DCASE25 SUBMISSION: ACOUSTIC SCENE CLASSIFICATION USING CNN-GRU MODEL WITHOUT KNOWLEDGE DISTILLATION

## Technical Report

Ee-Leng Tan[1], Jun Wei Yeow[1], Santi Peksi[1], Haowen Li[1], Ziyi Yang[1], Woon-Seng Gan[1]

[1] Smart Nation TRANS Lab, Nanyang Technological University,
50 Nanyang Avenue, Singapore 639798
etanel@ntu.edu.sg, junwei004@e.ntu.edu.sg, speksi@ntu.edu.sg, haowen.li@ntu.edu.sg,
ziyi016@e.ntu.edu.sg, ewsgan@ntu.edu.sg

## ABSTRACT

In this technical report, we present the SNTL-NTU team's Task 1 submission for the Low-Complexity Acoustic Scene Classification of the Detection and Classification of Acoustic Scenes and Events (DCASE) 2025 challenge [1]. This submission departs from the typical application of knowledge distillation from a teacher to a student model, aiming to achieve high performance with limited complexity. The proposed model is based on a CNN-GRU model and is trained solely using the TAU Urban Acoustic Scene 2022 Mobile development dataset [2], without utilizing any external datasets, except for MicIRP [3], which is used for device impulse response (DIR) augmentation. The proposed model has a memory usage of 114.2 KB and requires 10.9M multiply-and-accumulate (MAC) operations. Using the development dataset, the proposed model achieved an accuracy of 60.25%.

## 1. INTRODUCTION

In Task 1 of the DCASE Challenge 2025, acoustic scene classification (ASC) is employed to classify 10 acoustic scenes from 12 cities based on 1-second audio samples. To align ASC with the performance of typical edge devices, Task 1 [1] of the DCASE Challenge 2023 has imposed the following system complexity constraints:

- Maximum memory allowance: 128 KB
- Maximum number of MACs per inference: 30 MMAC

Numerous CNN models distilled from large pretrained teachers have dominated the submissions in DCASE Task 1 since these networks achieved the highest accuracies with the TAU Urban Acoustic Scene 2022 Mobile dataset [2] since 2022. Along with the knowledge distillation (KD), augmentation techniques were extensively applied to enhance the model's generalizability to unseen devices and the variability of the audio samples obtained for the same scenes from different cities. To further reduce the model size, post-training quantization and pruning were applied to the weights and parameters of models.

The proposed model is a shallow CNN network combining depthwise (DW) convolution, channel shuffle (CS), squeeze-and-excite (SE), and gated recurrent unit (GRU) to effectively perform ASC. The MAC of the proposed model is 10.9M and 16-bit precision is used to quantize the proposed model since the number of parameters is less than 64K.

The remaining sections of this report are organized as follows. In Section 2, the input features, augmentation techniques used, and proposed model are discussed. Section 3 presents the results of our submissions based on the various splits of the development dataset. This report is concluded in Section 4.

## 2. PROPOSED SYSTEM

### 2.1. Preprocessing

The TAU urban acoustic scene 2022 mobile dataset contains recordings of 10 acoustic scenes in 12 European cities. These recordings are captured using four devices and synthetic data for 11 devices was generated using the recordings. Each 1 sec audio sample is captured with a sampling frequency of 44.1 kHz sampling rate and encoded at 24-bit resolution.

For the input feature, a 256 bin mel-spectrogram is calculated using the short time Fourier transform (STFT) having a window length of 0.18 sec and 17% overlap. This configuration produces 33 time frames and an input feature shape of [256 × 33] for each audio sample.

The input features and the down-sampling of the audio samples are computed and performed using Librosa [3], respectively. The sampling rate 44.1kHz is selected based on the averaged mel spectrograms of the 10 acoustic scenes, as shown in Fig. 1. The acoustic scenes of the airport, park, shopping mall, street pedestrian, and tram are narrower in terms of bandwidth, while significant frequency components at 16kHz and above are observed in the remaining acoustic scenes.

### 2.2. Data Augmentation

Three augmentations using SpecAugment [4], Freq-MixStyle [5], and device impulse response (DIR) [6] are applied in proposed models to improve system robustness and performance.

SpecAugment typically consists of three types of augmentations: time-warping, frequency masking, and time-masking. For our model, frequency masking has proven particularly effective.

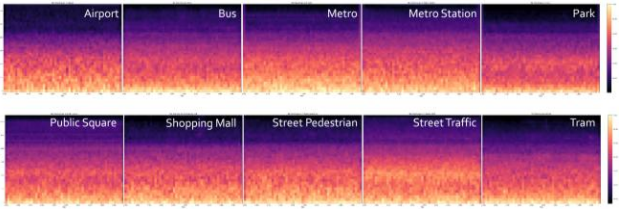

Fig. 1. Averaged mel-spectrograms of 10 acoustic scenes. Acoustic scenes of bus, metro, metro station, public square, and street traffic are found to span across a wider bandwidth.

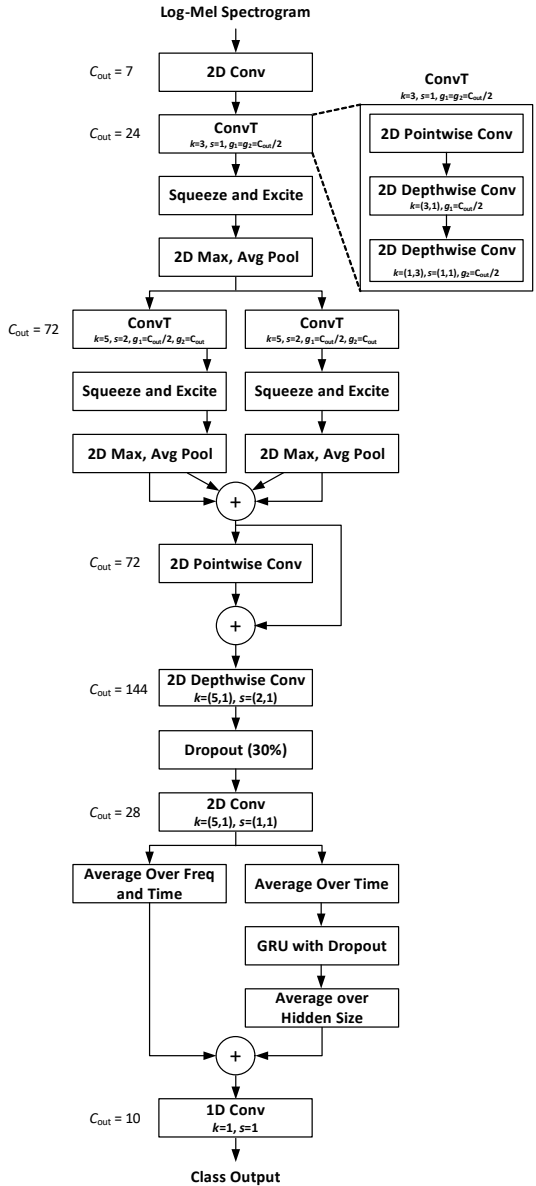

Fig. 2. Proposed GRU-CNN model. Output channels of convolution blocks are shown on the left. Input channel of model is one.

Freq-MixStyle extends MixStyle [7] to the frequency domain of the audio samples. By exposing the model to a variety of mixed spectral properties, the model can generalize to different acoustic environments and their variations. It has been shown that models trained with Freq-MixStyle exhibit better generalization to unseen conditions, enhancing domain invariance. Models trained with DIR augmentation can also better handle variations in recording devices, leading to improved accuracy with the development dataset. The only external dataset, MicIRP, is used in data augmentation.

### 2.3. Proposed GRU-CNN Model

The architecture of the submitted model is depicted in Fig. 2. It combines 2D and 1D convolutions, DW separable convolutions, SE blocks, GRU, and hybrid pooling (average and max pooling) strategies to extract rich spectral features from log-mel spectrograms.

To reduce computational complexity, standard 2D convolutions are replaced with DW separable and pointwise convolutions. Inspired by MobileNetV2, the architecture performs pointwise convolution before DW convolution, with channel expansion applied in both stages. The DW convolution is further decomposed into two spatially separated 1D convolutions—along the time and frequency axes—to more effectively capture distinct temporal and spectral patterns. Feature map downsampling is applied in the second 1D convolution to minimize information loss during resolution reduction. The pointwise and two 1D depthwise convolutions are encapsulated in the ConvT block as shown in Fig. 2. SE blocks are integrated alongside both max-pooling and average-pooling operations to enhance feature representation in the early and intermediate layers.

Unlike conventional sequence models that operate along the temporal dimension, the GRU in this architecture is configured to learn patterns across the frequency axis. Specifically, the input is structured with a feature size of $F$ (frequency bins) and a sequence length of $C$ (channels). This design is also a consequence of the large FFT size used to compute the log-mel spectrograms. As a result, the GRU produces an output of shape ($B$, $C$, $H$), where $B$ is the batch size and $H$ is the hidden size. These features are subsequently fused with those extracted by a parallel 1D convolutional layer, enabling complementary pattern learning across the log-mel spectrogram.

### 3. RESULTS AND SUBMISSION

For all splits, the proposed model was trained for 150 epochs with a batch size of 256 using the ADAM optimizer with a learning rate adjusted by the cosine schedule with ramp-up. The window length, hop length, FFT size, and number of mel bins are 8192, 1364, 8192, and 256, respectively. The results of the provided baseline and submitted models are summarized in Tables I and II, respectively. Compared to the baseline model, the proposed model demonstrates better classification across most classes.

### 4. CONCLUSIONS

In this technical report, we described the SNTL-NTU submissions to task 1 of the DCASE 2025 challenge. The proposed model is based on CNN and is trained solely on the TAU Urban Acoustic Scene 2022 Mobile development dataset.

Table I Class-Wise Accuracies of Baseline in Percentage

| Model | Airport | Bus | Metro | Metro Station | Park | Public Square | Shopping Mall | Street Pedestrian | Street Traffic | Tram | Macro Acc |
|---|---|---|---|---|---|---|---|---|---|---|---|
| General | 38.94 | 62.28 | 40.60 | 50.72 | 72.03 | 29.20 | 56.04 | 34.76 | 73.21 | 49.42 | 42.40 |
| Device | 44.43 | 64.81 | 43.87 | 48.22 | 72.75 | 32.04 | 53.14 | 34.43 | 74.10 | 51.08 | 51.89 |

Table II Device-Wise Accuracies of Baseline in Percentage

| Split | A | B | C | S1 | S2 | S3 | S4 | S5 | S6 | Acc |
|---|---|---|---|---|---|---|---|---|---|---|
| General | 62.80 | 52.87 | 54.23 | 48.52 | 47.29 | 52.86 | 48.14 | 47.23 | 42.60 | 50.72 |
| Device | 63.98 | 55.85 | 59.09 | 48.68 | 48.74 | 52.72 | 48.14 | 47.23 | 42.60 | 51.89 |

## 5. ACKNOWLEDGEMENT

This research is supported by the Ministry of Education, Singapore, under its Academic Research Fund Tier 2 (MOE-T2EP20221-0014).